\documentstyle[preprint,aps,prl]{revtex}
\begin{document}
\draft
\title{Large-Scale Flow and Spiral Core Instability in Rayleigh-B\'enard
       Convection}
\author{Igor Aranson$^1$, Michel Assenheimer$^2$,
        Victor Steinberg$^3$ and Lev S. Tsimring$^4$}
\address{
    $^1$Department of Physics, Bar Ilan University, 52900 Ramat Gan, Israel\\
    $^2$Laboratoire de Physique Statistique, Ecole Normale Sup\'erieure, 75231
        Paris cedex 05, France\\
    $^3$Department of Physics of Complex Systems, The Weizmann Institute of
        Science, 76100 Rehovot, Israel\\
    $^4$Institute for Nonlinear Science, University of California at San Diego,
        San Diego, CA 92093-0402}
\date{\today}
\maketitle
\begin{abstract}
The spiral core instability, observed in large aspect ratio Rayleigh-B\'enard
convection, is studied numerically in the framework of the Swift-Hohenberg
equation coupled to a large-scale flow. It is shown that the instability leads
to non-trivial core dynamics and is driven by the self-generated
vorticity. Moreover, the recently reported transition from spirals to hexagons
near  the core is shown to occur only in the presence of a non-variational
nonlinearity, and is triggered by the spiral core instability. 
Qualitative agreement between the simulations and the experiments is 
demonstrated.
\end{abstract}


One of the most intriguing and unexpected recent discoveries in natural pattern
formation is the observation of spatio-temporally disordered spiral and target
patterns\cite{assstei1,morris} in large aspect ratio Rayleigh-B\'enard
convection (RBC) in Boussinesq fluids, in a parameter range where
previously only
rolls were known to be stable\cite{busse}.
This regime is characterized by the spontaneous and continuous
emergence and annihilation of large extended spiral and target
patterns. Theoretically, these novel states were successfully reproduced
both by numerical simulations of the Swift-Hohenberg (SH) model coupled to a
self-consistent large-scale flow (see
Eqs.~\ref{SHE}-\ref{vel})\cite{xi,haken}, as well as by the integration of the
full thermally driven Navier-Stokes equations in the Boussinesq
approximation\cite{pesch}. It is currently postulated that
the large-scale flow is necessary for the spatio-temporal chaotic state with
many spirals and targets\cite{xi,haken,pesch,cross}.
In a first attempt to understand these states, Cross and Tu proposed a physical
mechanism based on wavenumber frustration, in which defects have an invasive
nature and create spirals and targets\cite{cross}.
In detailed experiments by Assenheimer and
Steinberg\cite{assthes,assstei}, and more recently by Plapp and
Bodenschatz\cite{boden}, a new instability of spiral cores of single- and
multi-armed spirals was observed. The striking feature of this instability is
that spiral cores oscillate periodically with a frequency considerably
higher than the frequency of the overall spiral
rotation\cite{boden}. For yet higher
supercriticality, a novel transition from spirals to hexagons was
found in which both up- and downflow
hexagons coexist\cite{assthes,assstei}. These hexagons often invade the
background RBC pattern, originating mainly from extended pattern cores.

In this Letter we present numerical simulations of the spiral core dynamics
performed in the framework of the SH model coupled to a
self-consistent large-scale flow. We show that this simple model
exhibits both the spiral core instability and the spiral-to-hexagon
transition, and that both are linked via the large-scale flow.
In the case of the spiral core instability, we demonstrate that the velocity
field generated by the spiral tip increases the local wavenumber and eventually
drives the tip into the skew-varicose unstable region. Phase slips then occur,
locally unwinding the spiral and returning the wavenumber into the stable
domain. This instability is found to have a well-defined threshold in Rayleigh
and Prandtl number space. Furthermore, the interpretation of the numerical
results is supported by the analysis of a similar but simpler problem: a
single-armed spiral in  an external velocity field created by a point vortex
located at the spiral core, in the limit of infinite Prandtl number when the
large-scale flow and the order parameter are decoupled. In addition, we
observed, at higher supercriticality and only in the presence of non-variational
nonlinear terms, a transition from spirals to hexagons. Both up- and downflow
hexagons are generated simultaneously near the spiral core.
In this scenario the core oscillations precede and trigger the transition
to the hexagonal state.

We considered the well-established model, which
describes RBC in a Boussinesq fluid rather successfully\cite{green},
\begin{eqnarray}
    \psi_t + ({\bf u} \cdot \nabla) \psi & = &
    \epsilon \psi - g \psi^3 + 3 (1 - g) (\nabla\psi)^2 \nabla^2 \psi -
    \nonumber \\
    & & - (1 + \nabla^2)^2 \psi \label{SHE} \\
    \Omega_t - \sigma (\nabla^2 - c^2) \Omega & = &
    g_m \hat{z} \cdot \nabla (\nabla^2 \psi) \times \nabla\psi
    \label{Omega} \\
    \Omega & = & \nabla\times {\bf u}
    \label{vel}
\end{eqnarray}
Here $\psi$ is the order parameter, ${\bf u}$ the horizontal velocity
field of the large-scale flow,  and $\Omega$ the vertical component of the vorticity.
The control parameter $\epsilon$ represents the reduced Rayleigh number, while
$\sigma$ characterizes the Prandtl number of the fluid. The parameter $g$
allows to more accurately reproduce the stability properties of convection
patterns, while $g_m$  characterizes the coupling strength between the order
parameter $\psi$ and the vorticity $\Omega$. The phenomenological parameter $c$
is introduced to describe the local dissipation of the
vorticity (e.g. due to friction at the bottom of the convection
cell)\cite{green,croshoh}. Thus, Eq.~\ref{SHE} describes the dynamics of
the order parameter $\psi$, while Eq.~\ref{Omega}, using the definition of the
vorticity (Eq.~\ref{vel}), represents the coupling of the large-scale flow field
${\bf u}$ and the order parameter. For $g = 1$ and $g_m = 0$
Eqs.~\ref{SHE}-\ref{vel} reduce to the Swift-Hohenberg equation (SHE).

We solved Eqs.~\ref{SHE}-\ref{vel} in a domain of
$256 \times 256$ mesh points using a pseudo-spectral method
based on the Fast Fourier Transform. The physical
domain size was typically restricted to $150 \times 150$.
Circular boundary conditions were enforced by ramping $\epsilon$ towards
negative values at distances $r > R_{max} = 55$. The computations were performed
on a parallel Cray J932 supercomputer, and verified on a $512 \times 512$ grid.

We started the simulations from initial conditions of the form 
$\psi= \cos( q r + n \theta)$, where $r$ and $\theta$ are the polar coordinates,
$q$ is the wavenumber, and $n=\pm1,\pm2,\ldots$ is the topological charge of the
spiral ($|n|$ is the number of spiral arms, while the sign corresponds to the
chirality). These initial conditions relax in about $10$ to $20$ horizontal
diffusion times to spirals. For sufficiently small values of the parameters
$\epsilon$ and $g_m$ (see Fig.\ \ref{figa}), the
spirals maintain a stable rigid rotation with an angular velocity depending
both on $g_m$ and $\epsilon$, as well as on the
topological charge $n$\cite{smallg}. Typical spatial distributions of the order
parameter and the vorticity field for one-armed spirals are shown,  e.g. in
Ref.~\cite{haken}. As can be seen, the spiral tip generates a highly localized
vorticity peak at the core. In the case of
$n$-armed spirals, $n$ identical vortices are created at the core.
In this regime spiral cores are stable and only experience a slow off-center
drift if the aspect ratio is not sufficiently large\cite{croshoh}.

For $\epsilon$ above some threshold, depending on $g_m$ (see Fig.\ \ref{figa}),
we observed a novel spiral core instability. In contrast to the off-center drift,
it persists even in an infinite system, since the unstable mode is localized near
the core. The main feature of the instability is that the spiral core oscillates
in the reference frame of the rotating spiral. The amplitude of the modulation,
caused by the core
oscillation,  increases from zero as the bifurcaton is crossed towards higher
values of
$\epsilon$ and $g_m$. The critical value of the control parameter $\epsilon$
depends on $g_m$ as well as on $n$.
Figure\ \ref{figa} shows the
bifurcation diagram as a function of $\epsilon$ and $g_m$ for $n =
1,2$\cite{smallg}. Figure\ \ref{figb} presents typical numerical as well as
experimental snapshots of the core oscillation for one-armed spirals\cite{rema}.
Similar behavior occurs for two-armed spirals.

The core oscillations can be illustrated by comparing the temporal
behavior of the order parameter at points near the 
core and at the rim. At the edge, the core
oscillations are negligible so that mainly the background rotation is sensed.
Figure~\ref{figc} shows plots for one-armed spirals for three
values of $\epsilon$. Figure~\ref{figc}(a) illustrates the rigid spiral rotation,
below the threshold for core oscillations. The central (solid
line) and peripheral points (dashed line) oscillate with the same frequency,
the unambiguous signature of the spiral's rigid rotation. Figures~\ref{figc}(b)
and \ref{figc}(c) show the spiral dynamics above criticality. Here,
the core oscillates at a much higher frequency than the peripheral point. The
core thus has a fast rotation in the framework of the spiral's overall rotation.

The corresponding experimental data are shown in Fig.~\ref{figd}.
Figure~\ref{figd}(a) shows the rigid body rotation of a one-armed
spiral below the onset of the core instability. Clearly, the oscillations near
and away from the core are phase-locked. Figure~\ref{figd}(b) presents the fast
oscillations of the core above the threshold of the instability. On this short
time scale the peripheral signal does not vary significantly and is not shown.
Similar dynamics were reported in Refs.~\cite{assthes,boden}.

Our numerical simulations of the full model (Eqs.~\ref{SHE}-\ref{vel}) suggest
that the vorticity generated by the spiral tip plays a major role in  the core
dynamics. Therefore, the origin of the core oscillations can easily be
understood - at least qualitatively - in the framework of the following
simplified model.  Consider a one-armed spiral solution of the form $A(r)
\cos\phi$, where $\phi = q r + \theta$, in the framework of the
SHE (Eq.~\ref{SHE}), coupled to an external velocity field
generated by a fixed point vortex with circulation $\Gamma$, placed at the
center of rotation. The sign of the vortex is chosen to correspond to the
self-generated vortex in the full model. Using the phase
approximation (i.e. neglecting the
variations of the amplitude $A(r)$), we obtain (see
Refs.~\cite{croshoh,newell}), that $\phi_t = \Gamma/r^2\partial_\theta\phi +
\ldots$ . Therefore, the local wavenumber increases linearly, $q_t \propto
\Gamma/r^2$. In other words, the external velocity field winds the spiral up
near the core. Eventually, the local wavenumber will be carried away from the
stable band and the skew-varicose (or Eckhaus) instability will be initiated.
This stage can no longer be described within the phase approximation. Abrupt 
phase jumps by
$2\pi$ (phase slips) consequently emerge and return the wavenumber back into the
stable region. This process then recurs, leading to quasi-periodic
oscillations.

We simulated the spiral dynamics using this
simplified model, with a velocity profile of the form ${\bf u} = \Gamma
r^{-1} \hat{\theta}$, and observed the abovementioned scenario. As the magnitude
of the circulation $\Gamma$ increases, a bifurcation similar to the one
described above occurs. At small $\Gamma$ a steady rotation persists,
while for $\Gamma > \Gamma_c$ the core starts to oscillate.
Interesting dynamics also
develop if the sign of the external circulation and
the spiral-generated vortex are opposite. Then, the local velocity
field unwinds the spiral, driving it into the zig-zag  unstable
region\cite{remb}. Because it is non-local, this instability cannot be saturated
by phase slips and  eventually destroys the spiral.
This might occur when a spiral is driven by the external vorticity
created by another oppositely charged spiral and could be relevant for the
persistence of the spiral turbulent state\cite{assstei1,morris}.

Recently Assenheimer and Steinberg reported a
transition from the spiral and target chaotic state to a state of up- and
downflow hexagons, as the supecriticality $\epsilon$ increased\cite{assstei}.
These hexagons started to develop and invade the system primarily from spiral 
and target cores and other defects. Dewel et al.\cite{dewel}  demonstrated that
in the framework of the SHE at large $\epsilon$, the coexistence and linear
stability of up- and downflow hexagons is caused by the excitation of a {\em
quasineutral} zero mode, which breaks the inversion  symmetry $\psi \rightarrow
-\psi$.  Despite their linear stability, these hexagons
are nonlinearly unstable because their free energy, in the framework of the pure
variational SHE ($g = 1$, $g_m = 0$), is higher than that of rolls.
As a result, nuclei of hexagons, immersed in rolls, ultimately shrink. 
Because the  generalized SHE (Eq.~\ref{SHE}-\ref{vel}) with either $g \ne 1$
and/or $g_m \ne 0$ is non-variational, a simple relative
stability analysis of rolls versus hexagons becomes impossible.

Numerics performed with a non-variational coefficient $1 - g = 0.25$,
relatively large coupling to the vorticity $g_m=10$ (customary for this model,
e.g. Ref.~\cite{croshoh}) and supercriticality above the threshold for the spiral
core instability, $\epsilon = 1.9$, show that hexagons indeed invade rolls. We
have observed the simultaneous nucleation of  up- and downflow hexagons at the
core which subsequently spread out (see Fig.~\ref{fige}).  Similar dynamics,
obtained experimentally, is shown in Fig.~\ref{figf}. However, one might
speculate that a local wavenumber change near the  core, rather than the
large-scale flow, causes the transition.  To compare, we performed simulations
for the same parameters but without a large-scale flow (i.e. $g = 0.75$, $g_m =
0$ and
$\epsilon = 1.9$).  In this case spots emerge from the side wall rather than
from  the core region. It is thus plausible to suggest that the large-scale 
flow, arising primarily near spiral cores and other defects, generates the 
zero mode of the order parameter\cite{zeromode}. Thus, the spiral core
instability initiates the nucleation of  hexagons. Since the $\psi
\rightarrow -\psi$ symmetry is still preserved, both up- and downflow hexagons
nucleate simultaneously.

Summarizing, we studied effects of large-scale flow
on the dynamics of a spiral core. Although our computations were performed
in the framework of a simplified phenomenological model,
two characteristic features of the dynamics were found also
observed experimentally: the spiral core instability and the spiral-to-hexagon
transition. The vorticity field generated at a spiral core when
$g_m\neq0$, plays a major role in the spiral oscillations. These
oscillations are
not observed in the variational model in which the coupling with the
vorticity mode is absent  (i.e. $g_m=0$). On the other hand, very similar
oscillations are observed in a model with a fixed vortex pinned at the
center of rotation of the spiral.
At higher supercriticality the vorticity drives the zero-mode of the order
parameter $\psi$ and prompts the hexagon formation. In that case, the core
oscillations initiate the transition to the hexagonal state.
Certainly, a more detailed analysis of a physically
more justified (but more complicated) model, similar to that of
Ref.~\cite{pesch}, is desirable.

I.A. was supported by the Raschi Foundation and
Israeli Science Foundation. M.A. benefitted from a Human Capital and
Mobility Fellowship. L.T. was supported by the U.S. Department of Energy and
acknowledges the hospitality of Bar Ilan University and The Weizmann
Institute of Science. Support by the Minerva Center for Nonlinear Physics of
Complex Systems and the Inter-Israeli Center for Supercomputing is also
acknowledged.

\begin{figure}
\caption{Stability diagram for one- (a) and two-armed (b) spirals for $c^2=2$,
$\sigma=1$ and $g=1$. As $\epsilon$ increases,  the spiral core becomes unstable
at the dashed line, regaining stability at the solid line as $\epsilon$
decreases.  
\label{figa}
}
\end{figure}
\begin{figure}
\caption{
Snapshots of periodic spiral core oscillations. Top row: simulations
with $\epsilon=1.45, g=0.9, g_m=27, c^2=2$ and $\sigma = 1$; time
delay between frames $5$, and integration domain radius $R=55$.
Bottom row: experiments with $\epsilon=2.88$
and $\sigma=4.5$.
}
\label{figb}
\end{figure}
\begin{figure}
\caption{
Order parameter $\psi$ at the core (solid) and periphery (dashed) of
a one-armed spiral with $g=1, g_m = 95, \sigma=5$ and $c^2=1$: (a) below
threshold ($\epsilon=1.1$); (b),(c) above threshold ($\epsilon=1.3$ and $1.35$).
}
\label{figc}
\end{figure}
\begin{figure}
\caption{
Shadowgraph intensity at the core (solid) and periphery
(dashed) of a one-armed spiral: (a) below
threshold ($\epsilon = 2.05, \sigma=4.5$) and (b)
above threshold ($\epsilon=2.88, \sigma=4.5$). In (b) only the fast core
signal is shown.
}
\label{figd}
\end{figure}
\begin{figure}
\caption{
Hexagon invasion from a spiral core obtained from the full model with
$\epsilon=1.9, g=0.75, g_m=10, c^2=2$ and $\sigma=1$. Snapshots taken at $t
= 10,110,470,650,340,2350$.
}
\label{fige}
\end{figure}
\begin{figure}
\caption{
Experimental hexagon nucleation at a spiral core with
$\epsilon = 3.19$, $\sigma = 4.5$ and time delay
between frames $\Delta t = 3.6$, $3.6$, $22.7$, $18.0$, $10.7$ $\tau_v$.
}
\label{figf}
\end{figure}

\begin{references}
\bibitem{assstei1}
M. Assenheimer and V. Steinberg, \prl {\bf 70}, 3888 (1993); Nature, {\bf 367},
345 (1994).
\bibitem{morris}
S. W. Moris, E. Bodenschatz, D.S. Cannel and G. Ahlers, \prl {\bf 71}, 2026
(1993).
\bibitem{busse} F.H. Busse, J. Fluid Mech. {\bf 30}, 625 (1967);
Rep. Prog. Phys. {\bf 41}, 1929 (1978).
\bibitem{xi}
H. Xi, J.D. Gunton, and J. Vi\~nals, \pre {\bf 47}, R2987 (1993)
\bibitem{haken}
M. Bestehorn, M. Frantz, R. Friedrich, and H. Haken, Phys. Lett. A.
{\bf 174}, 48 (1993).
\bibitem{pesch}
W. Decker, W. Pesch, and A. Weber, \prl {\bf 73},
648 (1994).
\bibitem{cross}
M.C. Cross and Y. Tu, \prl {\bf 75}, 834 (1995).
\bibitem{assthes}
M. Assenheimer, Ph. D. thesis, The Weizmann Institute of Science, 1994.
\bibitem{assstei}
M. Assenheimer and V. Steinberg, \prl {\bf 76}, 756 (1996); to be published.
\bibitem{boden}
B. B. Plapp and E. Bodenschatz, Core Dynamics of Multi-Armed Spirals in
Rayleigh-B\'enard Convection, submitted to Physica Scripta (1996).
\bibitem{green}
H. S. Greenside and M. C. Cross, Phys. Rev. A {\bf 31}, 2492 (1985).
\bibitem{croshoh}
M. Cross and P.C. Hohenberg, \rmp {\bf 65}, 851 (1993).
\bibitem{smallg} At even smaller $g_m$ spirals undergo a
spontaneous transition to targets via a core reconnection and expulsion of 
emerging dislocations off to the boundary (see Ref.~\protect\cite{assstei1}).
\bibitem{rema} The experimental data depicts the optical intensity
of the shadowgraph signal, related to the order parameter $\psi$.
\bibitem{newell}
M.C. Cross and A. C. Newell, Physica D {\bf 10}, 299 (1984).
\bibitem{remb} In this simplified model the Prandtl number is infinite.
Then, the first instability for $q < q_c$ is
zig-zag\protect\cite{croshoh}.
\bibitem{dewel} G. Dewel, S. M\'etens, M'F. Hilali, P. Borckmans, and
C.B. Price, \prl {\bf 74}, 4647, (1995).
\bibitem{zeromode} Large-scale flow and 
zero mode of the order parameter should not be equated. Physically, the latter
can be related to the long-wave variation of the vertical temperature
gradient from its nominal value. Note, while the zero mode has
different signs for up- and downflow hexagons, the vertical vorticity
has the same sign everywhere.
\end{references}
\end{document}